\documentclass[fleqn,usenatbib]{mnras}
\usepackage{newtxtext,newtxmath}
\usepackage[T1]{fontenc}
\DeclareRobustCommand{\VAN}[3]{#2}
\let\VANthebibliography\thebibliography
\def\thebibliography{\DeclareRobustCommand{\VAN}[3]{##3}\VANthebibliography}

\newcommand{\revised}[1]{{\textcolor{black}{#1} }}

\usepackage{graphicx}
\usepackage{amsmath}
\usepackage{xcolor}

\title[Using KNe to constrain the EoS]{Black Hole - Neutron Star mergers: using kilonovae to constrain the equation of state} 

\author[L. W. P. Mathias et al.]{
L. W. P. Mathias,$^{1,2}$\thanks{E-mail: lwpm20@outlook.com}
F. Di Clemente,$^{2,3}$
M. Bulla$^{2,3,4}$
and A. Drago$^{2,3}$
\\
$^{1}$Department of Physics, University of Bath, Claverton Down, Bath BA2 7AY, United Kingdom\\
$^{2}$Department of Physics and Earth Science, University of Ferrara, via Saragat 1, I-44122 Ferrara, Italy\\
$^{3}$INFN, Sezione di Ferrara, via Saragat 1, I-44122 Ferrara, Italy\\
$^{4}$INAF, Osservatorio Astronomico d’Abruzzo, via Mentore Maggini snc, 64100 Teramo, Italy\\
}

\date{Accepted 2023 December 16. Received 2023 December 15; in original form 2023 September 11}

\pubyear{2023}

\begin{document}
\label{firstpage}
\pagerange{\pageref{firstpage}--\pageref{lastpage}}
\maketitle

\begin{abstract}
The merging of a binary system involving two neutron stars (NSs), or a black hole (BH) and a NS, often results in the emission of an electromagnetic (EM) transient. One component of this EM transient is the epic explosion known as a kilonova (KN). The characteristics of the KN emission can be used to probe the equation of state (EoS) of NS matter responsible for its formation. We predict KN light curves from computationally simulated BH-NS mergers, by using the 3D radiative transfer code \texttt{POSSIS}.
We investigate two EoSs spanning most of the allowed range of the mass-radius diagram. We also consider a soft EoS compatible with the observational data within the so-called 2-families scenario in which hadronic stars coexist with strange stars.
Computed results show that the 2-families scenario, characterized by a soft EoS, should not produce a KN unless the mass of the binary components are small ($M_{\rm BH} \leq 6M_{\odot}$, $M_{\rm NS} \leq 1.4M_{\odot}$) and the BH is rapidly spinning ($\chi_{\rm BH} \geq 0.3$). In contrast, a strong KN signal potentially observable from future surveys (e.g. VRO/LSST) is produced in the 1-family scenario for a wider region of the parameter space, and even for non-rotating BHs ($\chi_{\rm BH} = 0$) when $M_{\rm BH} = 4M_{\odot}$ and $M_{\rm NS} = 1.2M_{\odot}$. We also provide a fit that allows for the calculation of the unbound mass from the observed KN magnitude, without running timely and costly radiative transfer simulations. Findings presented in this paper will be used to interpret light curves anticipated during the fourth observing run (O4), of the advanced LIGO, advanced Virgo and KAGRA interferometers and thus to constrain the EoS of NS matter. 

\end{abstract}

\begin{keywords}
equation of state - radiative transfer – methods: numerical – stars: neutron – black hole neutron star mergers - gravitational waves.
\end{keywords}

\section{Introduction}

Neutron stars (NSs) comprise the densest stable matter of any observable object known to date. They lie on the threshold between ultra-dense stellar matter and a black hole (BH). Uncovering the structure and nature of matter at this threshold by establishing a single unified equation of state (EoS) has proven challenging; it remains one of the great enigmas of modern astronomy. The 2017 discovery of a transient event identified as the electromagnetic (EM) counterpart of the gravitational wave (GW) produced by the double NS merger GW170817 marked year zero of the multi-messenger GW era \citep{Abbott2017}; it divulged the potential of such mergers to unveil a richer picture of the structure and nature of matter inside compact objects such as NSs. 

%GWs and their EM counterpart offer the advantage of constraining multiple properties (mass, radius and microphysics) of NSs at once, unlike alternative methods (such as X-ray and pulsar detections) that can each constrain only one \citep{Ozel:2016oaf}. 

The EM transient comprises several observable components: gamma ray bursts (GRBs) powered by relativistic jets, the associated GRB afterglow from the interaction of the jet with the interstellar medium, and KNe. KNe are produced when matter ejected by double NS and BH-NS mergers (i.e. merger ejecta) undergoes rapid neutron capture ($r$-process) nucleosynthesis, producing unstable neutron-rich elements \citep{Metzger2019}. The rapid beta decay of these heavy elements powers the epic explosion known as a KN. Of the aforementioned EM components, KNe’s approximately isotropic nature and relatively long emission timescales (days) make them the most attractive of the available candidates for observation of GW-detected mergers. Encoded within KN observables (light curves, spectra and polarisation) detectable from Earth is information on the physical conditions under which they were produced. EM transients and their associated KN will thus be key to placing further constraints on the EoS of NS matter.

As alluded to above, matter can be ejected during double NS and BH-NS mergers. The properties of this merger ejecta (e.g. mass and velocity) are dependent on both the binary properties of the merger system (masses and spins of the binary components) and the EoS of NS matter. The merger ejecta in turn determines the characteristics of the KN. Using this relation between the progenitor and the KN, the EoS  of NS matter can be probed if the binary properties of the merger system are known (these can be identified from the GW preceding the KN).

\begin{figure}
\includegraphics[width=\columnwidth]{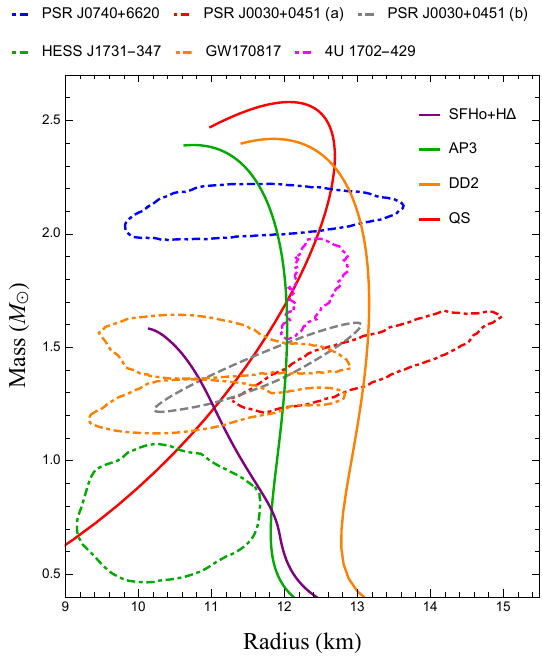}
     \caption{Mass-radius relation from a few EoSs (solid), see text. \revised{1$\sigma$} observational constraints (dot-dashed) for PSR J0740+6620  \citep{Miller2021}, PSR J0030+0451 \revised{(analysis \textit{a}) \citep{Riley2019}, PSR J0030+0451 (analysis \textit{b})}\citep{Vinciguerra:2023qxq}, HESS J1731-347 \citep{Doroshenko2022}, GW170817 \citep{AbbottTD2018}, 4U 1702-429 \citep{Nattila2017}.
    }
    \label{fig:M-R_figure}
\end{figure}

The 1-family versus 2-families scenario refers to two theories on how compact stars exist in nature. The 1-family scenario assumes the existence of only one family of compact stars. These compact stars could be NSs (made only of nucleons), hadronic stars (HSs, in which hyperons and/or various resonances appear above a threshold density) or hybrid stars (in which deconfined quarks are present above a certain density); and which it is, is currently unknown. The 1-family scenario assumes that matter inside compact stars is described by a unique EoS, and that the ground state of matter is hadronic, namely $^{56}$Fe. If the 1-family scenario holds true in nature, the EoS must be relatively stiff in order to satisfy observational constraints imposed by analysis of the pulsar PSR J0740+6620 studied by NICER (The Neutron Star Interior Compositions ExploreR, \citet{Raaijmakers2021}, \citet{Riley2021}, \citet{Miller2021}) and by the existence of massive compact stars as PSR J09562-0607 \citep{Romani:2022jhd}, see Fig.~\ref{fig:M-R_figure}. A stiff EoS describes a star with a large increase in pressure for a given increase in density. This results in stars with relatively large radii and low compactness that are expected to eject more material in a BH-NS merger, due to their large tidal deformability, resulting in a brighter KN.

The 2-families scenario refers to the possible coexistence of two families of compact stars in nature: HSs and strange quark stars (QSs). It describes a phenomenology where hadronic matter is metastable, whilst strange quark matter is absolutely stable; this naturally occurs when the Bodmer-Witten hypothesis is considered \citep{Bodmer1971,Witten:1984rs}.

The scenario was developed to justify the possible existence of compact stars with radii smaller than those compatible with the 1-family scenario \citep{Drago:2013fsa}.
\revised{Indeed, it has been shown that the radius of a 1.4 $M_\odot$ star cannot be smaller than about (11-11.5) km in the absence of a strong quark deconfinement phase transition \citep{Most2018,Coughlin:2019,Radice:2018ozg,Kashyap:2021wzs}}. Within the 2-families scenario this limit can be violated: HSs can reach significantly smaller radii (but not very large masses), while the most massive compact stars correspond to QSs. HSs here are described by a soft EoS, due to the inexorable formation of hyperons and delta resonances at densities higher than those seen in the 1-family scenario \citep{Drago2014_2}. This results in HSs of smaller radii, and thus greater compactness than in the 1-family scenario.
The 2-families scenario can be discounted if it is shown that all compact stars of mass $\sim 1.4 M_\odot$ have radii $\gtrsim11.5\rm km$  \citep{DiClemente2022}. 
Notice that, although QSs can reach very large masses \citep{Bombaci2021}, numerical simulations in  \citet{Kluzniak2002} suggest that they do not produce ejecta (because of the sticking properties of strange quark matter \citet{Bauswein:2009im}) and in turn a KN following a merger. Therefore, within the 2-families scenario, only mergers involving HSs and a BH are investigated in this paper. More extensive numerical simulations will be needed in order to confirm that no matter is ejected in a BH-QS merger.

In this paper, two EoSs from the 1-family scenario, and a soft Eos from the 2-families scenario are investigated; these EoSs were selected based on their compatibility with recent observational constraints, including those placed by NICER and GW170817, see Fig.~\ref{fig:M-R_figure}.
The two EoSs of the 1-family scenario span a significant fraction of the available range in the mass-radius plane, if only one family is considered. The soft EoS of the 2-families scenario is compatible with small radii as the ones obtained from many X-ray sources (not displayed in Fig.~\ref{fig:M-R_figure}) and discussed e.g. in fig. 4 of \citet{Ozel:2016oaf}. Interestingly, a very recent re-analysis of PSR J0030+0451 \citep{Vinciguerra:2023qxq} concluded that object can have either a mass of $1.70 ^{+0.18}_{-0.19} M _\odot$ and a very large radius $14.44^{+0.88}_{-1.05}$ km (probably not compatible with other astrophysical data and not displayed in Fig.~\ref{fig:M-R_figure}) or a mass $\sim 1.4 M_\odot$ and a rather small radius, compatible with the two most soft hadronic EoSs (AP3 and SFHo) scenario and also with a QS.

By looking at Fig.~\ref{fig:M-R_figure} one is tempted to conclude that all the available constraints on masses and radii can be satisfied by the QS branch, including the recent analysis of HESS J1731-347, indicating a subsolar mass and a small radius. Indeed we have shown that QSs can account for the existence of compact stars having very small \citep{DiClemente2022_b} or very large masses \citep{Bombaci2021}. On the other hand, it is unlikely that all compact stars are QSs: it is well known that magnetar oscillations pose challenges to QSs \citep{Watts:2006hk}. Also, the analysis of the energy released by the SN1987A indicates a binding energy perfectly compatible with that of a NS \citep{Pagliaroli:2008ur}. Instead, in order to satisfy the limit on the mass of HESS J1731-347, one needs to use the significantly larger value of the binding energy of a QS \citep{DiClemente2022_b}.

In this work, KN light curves are predicted from BH-NS merger simulations to aid in constraining the EoS of compact star matter. A similar analysis was carried out in  \citet{DiClemente2022}; we build on this analysis by improving the modelization of the KN using the 3D radiative transfer code \texttt{POSSIS} \citep{Bulla2019,Bulla2023}.

\begin{figure*}
	\includegraphics[width=0.95\textwidth]{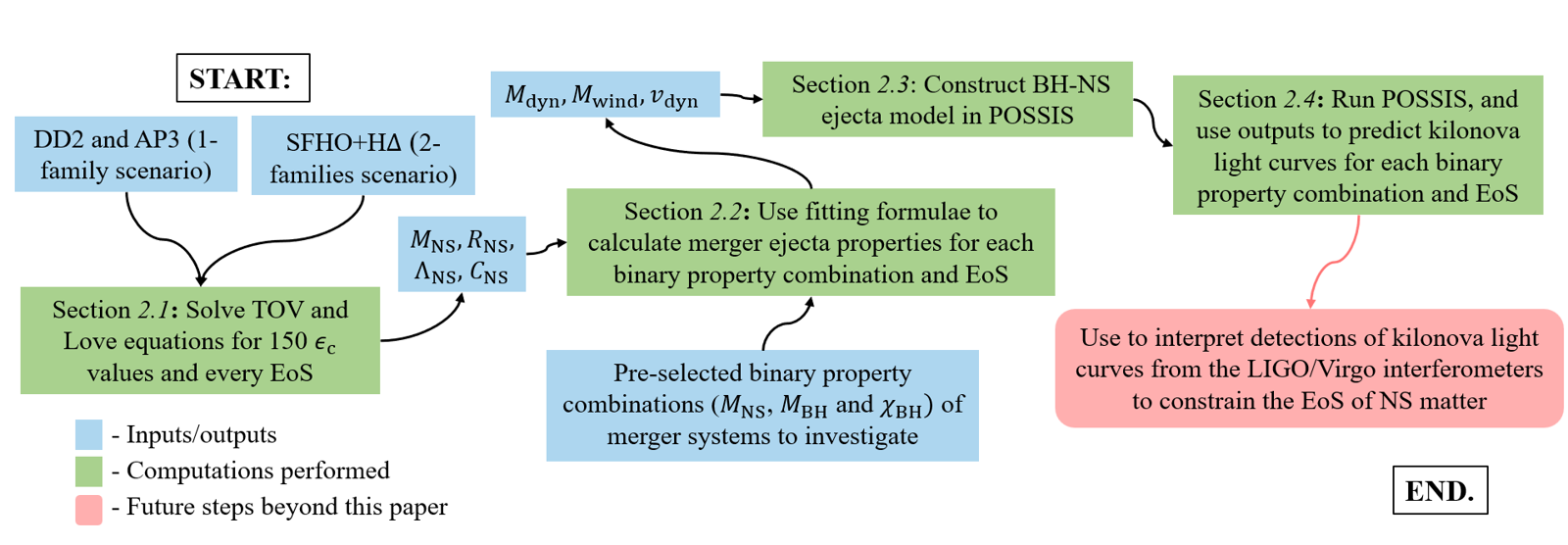}
    \caption{ Summary of method steps. Simulation steps (in green) computations performed in each of the methods sections (\ref{sec:2.1}, \ref{sec:2.2}, \ref{sec:2.3} and \ref{sec:2.4}); the inputs and outputs of these computations are shown in blue. Future steps beyond the scope of this paper are shown in pink. 
    }
    \label{fig:Methods steps}
\end{figure*}
\section{Methods}
In this section, we present the method steps taken to predict KN light curves for various BH-NS merger systems and EoSs. Three EoSs were investigated; two stiff EoSs (DD2, \citet{Hempel2010}, and AP3, \citet{Akmal1998}, both hadronic) from the 1-family scenario, and one soft EoS (SFHo+H$\Delta$, \citet{Drago2014_2}, also hadronic) from the 2-families scenario. KN light curves were predicted from these EoSs for various BH-NS merger systems by implementing the methods described in this section; a diagram of these methods steps can be found in Fig.~\ref{fig:Methods steps}.

\subsection{Calculating the compactness and tidal deformability from an EoS}
\label{sec:2.1}
The Tolman-Oppenheimer-Volkoff (TOV) equation \citep{Oppenheimer1939} was solved using the chosen EoSs to calculate the mass ($M_{\rm NS}$) and radius ($R_{\rm NS}$) of a NS for a given range of central energy densities ($\epsilon_{\rm c}$); $M_{\rm NS}$ and $R_{\rm NS}$ were calculated for 150 linearly spaced $\epsilon_c$ values between $1.4\times10^{14}$ and $2.5\times10^{15}$g cm$^{-3}$.  The $\epsilon_{\rm c}$ values were chosen so that the calculations for $M_{\rm NS}$ and $R_{\rm NS}$ covered their full range of possible values within current observational constraints \citep{Nattila2017,AbbottTD2018,Riley2019,Miller2021,Doroshenko2022}. We note that the EoSs investigated are only marginally compatible (90$\%$ confidence interval) with \citet{Doroshenko2022}; see \citet{DiClemente2022_b} for other possible astrophysical paths to form such a compact object that better satisfies these constraints. Calculating $M_{\rm NS}$ and $R_{\rm NS}$ across all parameter space allowed for accurate interpolation of the $M_{\rm NS}$-$R_{\rm NS}$ relation of NSs, and thus compactness, $C_{\rm NS}$=$M_{\rm NS}$/$R_{\rm NS}$, across all parameter space. Fig.~\ref{fig:M-R_figure} shows the $M_{\rm NS}$-$R_{\rm NS}$ for each EoS, together with current observational constraints. 

The tidal deformability, $\Lambda_{\rm NS}$, was calculated using equation (\ref{tidal}) below and those in  \citet{Hinderer2008} using results obtained from the TOV equation for each of the EoSs discussed above. $\Lambda_{\rm NS}$ is a dimensionless quantity defined as

\begin{equation}
\label{tidal}
    \Lambda_{\rm NS}=\frac{2}{3}k_2C_{\rm NS}^{-5}
\end{equation} where $k_2$ is the Love number \citep{Hinderer2008};  $k_2$ is another dimensionless quantity that describes the rigidity of a star. 

\subsection{Calculating merger ejecta properties}
\label{sec:2.2}

During a merger, matter is ejected if the NS is tidally disrupted. This occurs when the separation of the binary components ($d_{\rm tidal}$), at which tidal forces become strong enough to disrupt the star, is greater than the innermost stable circular orbit of the BH ($R_{\rm ISCO}$): $d_{\rm tidal}>R_{\rm ISCO}$. $d_{\rm tidal}$ is approximated by
\begin{equation}
\label{d_tidal}
    d_{\rm tidal} \approxeq R_{\rm NS} \left(\frac{3M_{\rm BH}}{M_{\rm NS}} \right) ^ {1/3}
\end{equation} where $M_{\rm NS}$, $M_{\rm BH}$, $R_{\rm NS}$ are the NS mass, BH mass and NS radius respectively, and $R_{\rm ISCO}$ is solely a function of black hole spin ($\chi_{\rm BH}$) and is defined in equations (\ref{Risco}-\ref{Z2}) \citep{Foucart2012}. In cases where this condition is met, merger ejecta produced by BH-NS binary mergers can broadly be categorised into two components: dynamical ejecta and those ejected from the bound disk \citep{Nakar2020}. The calculation of some key properties of these components, which directly determine the brightness of the resultant KN,  is discussed below.

In our analysis, 36 different BH-NS merger systems were investigated, each with a unique combination of binary properties, i.e. $M_{\rm NS}$, $M_{\rm BH}$ and $\chi_{\rm BH}$; details of their values can be found in Table~\ref{tab:table_1}. Merger ejecta properties were calculated for every merger system for each EoS in turn, resulting in 108 calculations in total (36 merger systems $\times3$ EoS). Calculations of the ejecta properties involved the use of fitting formulae from the semi-analytical models \citep{Foucart2018, Barbieri2020, Kawaguchi2016}. As well as being functions of  $M_{\rm NS}$, $M_{\rm BH}$ and $\chi_{\rm BH}$, these fitting formulae also require $C_{\rm NS}$ and $\Lambda_{\rm NS}$, which are computed as described in Section~\ref{sec:2.1}. The values of the binary properties were chosen based on observational constraints \citep{AbbottTD2018,Raaijmakers2021,AbbottR2021} and previous research on BH-NS mergers in the 1-family versus 2-families scenario, to ensure that many of the combinations would produce merger ejecta and thus a KN \citep{DiClemente2022}. 

% Table 1
\begin{table}
	\centering
	\caption{Summary of the merger systems investigated: 36 merger systems each with a unique combination of the $M_{\rm NS}$, $M_{\rm BH}$ and $\chi_{\rm BH}$ values listed here (4$M_{\rm NS}$ $\times3M_{\rm BH}$ $\times3\chi_{\rm BH}$). Each merger system was investigated for every EoS in turn.}
	\label{tab:table_1}
 
	\begin{tabular}{lc} 
		
		$\textbf{EoS:}$ &  DD2, AP3, SFHo+H$\Delta$ \\ 
        $\textbf{$M_{\rm NS}$ ($M_{\odot}$):}$ & 1.2, 1.4, 1.6, 1.8 \\
        $\textbf{$M_{\rm BH}$ ($M_{\odot}$):}$ & 4, 6, 8 \\
        $\textbf{$\chi_{\rm BH}$:}$ & 0, 0.3, 0.6 \\
		
	\end{tabular}
\end{table}

The ejecta properties calculated from the fitting formulae were the mass of the dynamical ejecta ($M_{\rm dyn}$), wind ejecta ($M_{\rm wind}$) and the average velocity of the dynamical ejecta ($v_{\rm dyn}$); this ejecta, all gravitationally unbound from the merger remnant, directly determines the brightness of the resultant KN. To determine these properties, several other quantities must first be determined. 
%All masses discussed here are in units of solar masses. 
The total ejected mass, $M_{\rm tot}$, the sum of $M_{\rm dyn}$ and the mass of the bound disk ($M_{\rm disk}$), is defined as
\begin{equation}
\label{Mtot}
    M_{\rm tot} = M_{\rm NS}^{\rm b} \left[{\rm max} \left\{\alpha\frac{1-2\rho}{\eta^{1/3}} - \beta\tilde{R}_{\rm ISCO}\frac{\rho}{\eta} + \gamma, 0 \right\}\right]^{\delta}
\end{equation} where $\alpha$, $\beta$, $\gamma$ and $\delta$ are fitting parameters defined in  \citet{Foucart2018}, and $\rho = (15\Lambda_{\rm NS})^{-1/5}$ is a function of the NSs $\Lambda_{\rm NS}$. The symmetric mass ratio
\begin{equation}
\label{symmetricmassratio}
    \eta = \frac{q}{(1+q)^2},
\end{equation} is a function of the binary mass component $q=M_{\rm NS}/M_{\rm BH}$. $\tilde{R}_{\rm ISCO} = R_{\rm ISCO}/M_{\rm NS}$ is the dimensionless normalised radius of the innermost stable circular orbit, and is defined as
\begin{equation}
\label{Risco}
    \tilde{R}_{\rm ISCO}(\chi) = 3 + Z_2 - {\rm sgn} \left( \sqrt{(3-Z_1) (3+Z_1+2Z_2} \right)
\end{equation} where
\begin{equation}
\label{Z1}
    Z_1 = 1 + \left( 1 - \chi^2 \right)^{1/3} \left[ \left( 1 + \chi \right)^{1/3} + \left( 1 - \chi \right)^{1/3}\right]
\end{equation} and
\begin{equation}
\label{Z2}
    Z_2 = \left( 3\chi^2 + Z_1^2 \right)^{1/2} .
\end{equation} The total baryonic mass of the NS, $M_{\rm NS}^b$, is the mass contained within baryonic matter, and is defined as
\begin{equation}
\label{Mbaryon}
    M_{\rm NS}^{\rm b} = M_{\rm NS} \left( 1 + \frac{0.6C_{\rm NS}}{1 - 0.5C_{\rm NS}} \right)
\end{equation} Equations (\ref{Mtot}-\ref{Mbaryon}), in addition to equations (\ref{Mdyn}-\ref{vdyn}) were solved to calculate $M_{\rm dyn}$, $M_{\rm wind}$ and $v_{\rm dyn}$. $M_{\rm dyn}$ was calculated from 
\begin{multline}
\label{Mdyn}
    M_{\rm dyn} = M_{\rm NS}^{\rm b} \left[ {\rm max} \left\{ \frac{a_1q^{-n_1}(1-2C_{\rm NS})}{C_{\rm NS}} - \right. \right. \\ 
    \left. \left. a_2q^{-n_2}\tilde{R}_{\rm ISCO}
    + a_3\left( 1 - \frac{M_{\rm NS}}{M_{\rm NS}^{\rm b}} \right) + a_4, 0 \right\} \right]
\end{multline} 
where $a_1$, $a_2$, $a_3$, $a_4$, $n_1$ and $n_2$ are fitting parameters defined in  \citet{Kawaguchi2016}. In equations (\ref{Mtot}) and (\ref{Mdyn}), $\tilde{R}_{\rm ISCO}$ was calculated using the parallel spin component $\chi_{\rm BH, \parallel}= \chi_{\rm BH} \cos({\rm l_{tilt}})$. Here, ${\rm l_{tilt}}$ is the angle between $\chi_{\rm BH}$ and the total angular momentum; it was assumed to be 0 for simplicity. However,  \citet{Foucart2020} suggests that using $\chi_{\rm BH}$ in place of $\chi_{\rm BH, \parallel}$ in the fitting formulae provides equally accurate predictions. Numerical simulations of BH-NS mergers in  \citet{Foucart2019} suggest that $M_{\rm dyn}$ cannot exceed  
\begin{equation}
\label{Mdynmax}
    M_{\rm dyn, max} = f_1M_{\rm tot}
\end{equation} where $f_1$ is a fraction that varies based on the mass ratio of the binary components; $f_1=0.35$ was assumed based on the values of $M_{\rm NS}$ and $M_{\rm BH}$ used in this work. 

\revised{We note that the fitting formulae from \cite{Kawaguchi2016} and \cite{Foucart2018} are valid within a range of $C_{\rm NS}$ values that encompasses those obtained with the SFHo+H$\Delta$ EoS in the two family scenario}\footnote{\revised{Below we find that material is ejected for one system with a NS with mass of 1.4\,$M_\odot$, a value that is higher than 1.35\,$M_\odot$, the maximum value allowed by the range of $C_{\rm NS}$ in \cite{Foucart2018}. However, we verified that results are only slightly affected by adopting 1.35 instead of 1.4\,$M_\odot$ , i.e. no system produces a kilonova except the same one, with ejecta masses and kilonova light curves that are similar ($\Delta M\sim0.01\,M_\odot$ and $\Delta {\rm mag}\lesssim0.5$)}.}.
\revised{It is important to note that microphysics, e.g. the presence of hyperons and delta resonances in the EoS SFHo+H$\Delta$, is crucial in determining the macroscopic parameters describing a compact star, as $\Lambda_{\rm NS}$, $C_{\rm NS}$ and also the threshold mass in a NS-NS merger. Ejecta properties, from numerical simulations of mergers, have been shown, at least in the case of NS-NS systems, to depend only on these three parameters \citep{Hotokezaka:2011dh,DePietri:2019khb}. Once they are given, the amount of mass dynamically ejected and the mass of the disk can be estimated from the fitting formulae. On the other hand, microphysics can still play a crucial role when transport properties as viscosity and thermal conductivity are tested in the numerical simulations, a problem which has not yet been systematically investigated.}

$M_{\rm disk}$ was calculated from the difference between $M_{\rm tot}$ and $M_{\rm dyn}$,
\begin{equation}
\label{Mdisk}
    M_{\rm disk} = {\rm max} \left\{ M_{\rm tot} - M_{\rm dyn}, 0 \right\}.
\end{equation} Since KNe originate from material that is gravitationally unbound from the merger remnant, the bound $M_{\rm disk}$ does not directly contribute. However, a fraction of this disk can be ejected as a wind via different processes \citep{Nakar2020}. The unbound wind mass was calculated from
\begin{equation}
\label{Mwind}
    M_{\rm wind} = f_2 M_{\rm disk}
\end{equation} where $f_2$ is a fraction between $0.2$ and $0.4$ \citep{Nakar2020}. There is uncertainty surrounding the true value of $f_2$ due to multiple complex mechanisms simultaneously determining the evolution of disk ejecta; $f_2=0.4$ was used following  \citet{Kawaguchi2020}, where BH-NS merger simulations similar to those undertaken here (see section \ref{sec:2.3}) were executed.

The average velocity of the dynamical mass is a function of the mass ratio, and was calculated from
\begin{equation}
\label{vdyn}
    v_{\rm dyn} = \left( 0.01533q^{-1} + 0.1907 \right) c~~~,
\end{equation}
where $c$ is the speed of light. Of the initial 36 merger systems, 23 produced ejecta and thus a KN for at least 1 of the 3 EoSs in question; stiffer EoSs (from the 1-family scenario) were more likely to produce a KN than a softer EoS (from the 2-families scenario). Across all 3 EoSs, 37 of the 108 possible combinations of binary properties and EoS (4$M_{\rm NS}$ $\times3M_{\rm BH}$ $\times3\chi_{\rm BH}$ $\times3 {\rm EoS}$) produced ejecta; 20 for DD2, 13 for AP3, and 4 for SFHo+H$\Delta$. Calculated values of $M_{\rm dyn}$, $M_{\rm wind}$ and $v_{\rm dyn}$ for each of these 37 merger systems can be found in Table~\ref{tab:table_2}.

% Table 2
\begin{table}
	\centering
	\caption{Calculated values of$M_{\rm dyn}$, $M_{\rm wind}$ and $v_{\rm dyn}$ for each of the 37 merger systems that produced ejecta. $M_{\rm NS}$, $M_{\rm BH}$, $M_{\rm dyn}$ and $M_{\rm wind}$ are expressed in units of solar masses, whilst $v_{\rm dyn}$ is expressed in units of the speed of light, $c$.}
	\label{tab:table_2}
	\begin{tabular}{lcccr} 
			
        $\textbf{EoS}$ &
            \begin{tabular}{@{}c@{}}\textbf{Merger system:} \\  
                $M_{\rm NS}$, $M_{\rm BH}$, $\chi_{\rm BH}$\end{tabular} &    
		$M_{\rm dyn}$ & $v_{\rm dyn}$ & $M_{\rm wind}$   \\

        \hline

        DD2 & 1.2, 4, 0 & 0.01906 & 0.199 & 0.01416 \\
        DD2 & 1.2, 4, 0.3 & 0.04509 & 0.199 & 0.03350 \\
        DD2 & 1.2, 4, 0.6 & 0.07836 & 0.199 & 0.06305 \\
        DD2 & 1.2, 6, 0 & 0.00015 & 0.224 & 0.00011 \\
        DD2 & 1.2, 6, 0.3 & 0.01917 & 0.224 & 0.01424 \\
        DD2 & 1.2, 6, 0.6 & 0.05892 & 0.224 & 0.04377 \\
        DD2 & 1.2, 8, 0.3 & 0.00152 & 0.249 & 0.00113 \\
        DD2 & 1.2, 8, 0.6 & 0.03616 & 0.249 & 0.02686 \\
        DD2 & 1.4, 4, 0 & 0.00352 & 0.192 & 0.00261 \\
        DD2 & 1.4, 4, 0.3 & 0.02552 & 0.192 & 0.02048 \\
        DD2 & 1.4, 4, 0.6 & 0.04216 & 0.192 & 0.06050 \\
        DD2 & 1.4, 6, 0.3 & 0.00421 & 0.213 & 0.00313 \\
        DD2 & 1.4, 6, 0.6 & 0.04177 & 0.213 & 0.03103 \\
        DD2 & 1.4, 8, 0.6 & 0.01877 & 0.234 & 0.01394 \\
        DD2 & 1.6, 4, 0.3 & 0.00527 & 0.187 & 0.00588 \\
        DD2 & 1.6, 4, 0.6 & 0.02137 & 0.187 & 0.04400 \\
        DD2 & 1.6, 6, 0.6 & 0.02052 & 0.205 & 0.01524 \\
        DD2 & 1.6, 8, 0.6 & 0.00229 & 0.224 & 0.00170 \\
        DD2 & 1.8, 4, 0.6 & 0.01647 & 0.182 & 0.01651 \\
        DD2 & 1.8, 6, 0.6 & 0.00158 & 0.199 & 0.00117 \\
        AP3 & 1.2, 4, 0 & 0.00464 & 0.199 & 0.00345 \\
        AP3 & 1.2, 4, 0.3 & 0.02639 & 0.199 & 0.01960 \\
        AP3 & 1.2, 4, 0.6 & 0.05297 & 0.199 & 0.05084 \\
        AP3 & 1.2, 6, 0.3 & 0.00401 & 0.224 & 0.00298 \\
        AP3 & 1.2, 6, 0.6 & 0.03799 & 0.224 & 0.02822 \\
        AP3 & 1.2, 8, 0.6 & 0.01617 & 0.249 & 0.01201 \\
        AP3 & 1.4, 4, 0.3 & 0.00752 & 0.192 & 0.00559 \\
        AP3 & 1.4, 4, 0.6 & 0.02564 & 0.192 & 0.03931 \\
        AP3 & 1.4, 6, 0.6 & 0.01823 & 0.213 & 0.01354 \\
        AP3 & 1.4, 8, 0.6 & 0.00125 & 0.234 & 0.00093 \\
        AP3 & 1.6, 4, 0.6 & 0.01555 & 0.187 & 0.01605 \\
        AP3 & 1.6, 6, 0.6 & 0.00099 & 0.205 & 0.00074 \\
        AP3 & 1.8, 4, 0.6 & 0.00012 & 0.182 & 0.00009 \\
        SFHo+H$\Delta$ & 1.2, 4, 0.3 & 0.00577 & 0.199 & 0.00429 \\
        SFHo+H$\Delta$ & 1.2, 4, 0.6 & 0.03542 & 0.199 & 0.02828 \\
        SFHo+H$\Delta$ & 1.2, 6, 0.6 & 0.01317 & 0.224 & 0.00979\\
        SFHo+H$\Delta$ & 1.4, 4, 0.6 & 0.00735 & 0.192 & 0.00546

	\end{tabular}
\end{table}

\subsection{Ejecta model}
\label{sec:2.3}

\begin{figure}
	\includegraphics[width=\columnwidth]{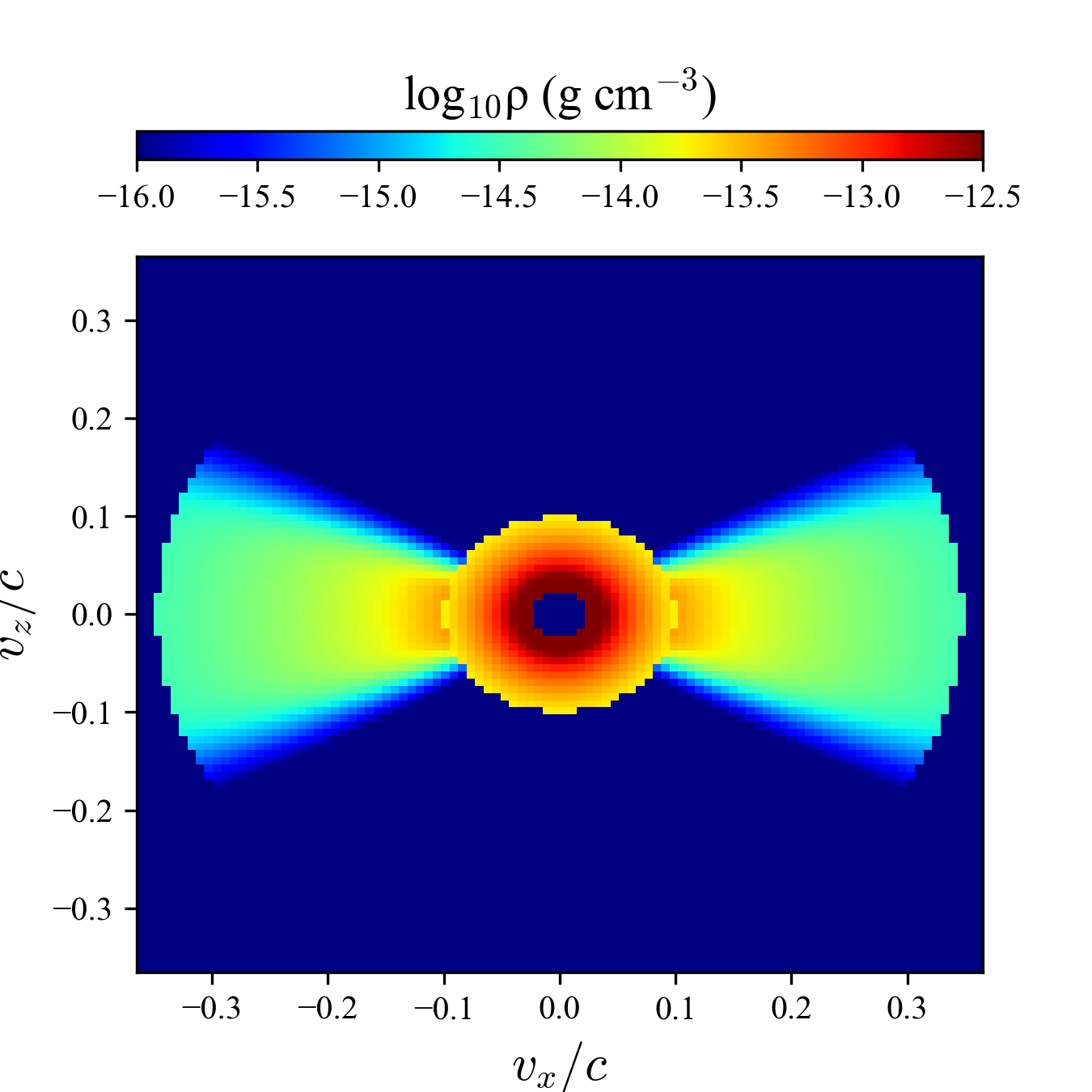}
    \caption{ 2D representation of the 3D density distribution grid used in \texttt{POSSIS}: central circle showing the $M_{\rm wind}$ ($0.025c-0.1c$) and lobes about the merger plane showing $M_{\rm dyn}$ ($0.1c-v_{\rm dyn, max}$). This grid illustrates the density distribution from the binary property combination $M_{\rm NS} = 1.2M_{\odot}$, $M_{\rm BH} = 6M_{\odot}$ and $\chi_{\rm BH}=0.3$ for the EoS AP3; values of the densities in each cell vary for different combinations of binary properties and EoS, but the distribution remains the same.
    }
    \label{fig:density}
\end{figure}

Calculated ejecta properties ($M_{\rm dyn}$, $M_{\rm wind}$ and $v_{\rm dyn}$) were used as input to the 3D radiative transfer (RT) code \texttt{POSSIS} (POlarization Spectral Synthesis In Supernovae, \citep{Bulla2019,Bulla2023}). \texttt{POSSIS} models the evolution of EM radiation through space; from its creation at an event such as a merger, its interaction with matter via absorption and scattering processes, to an observer on Earth. It was used in this work to predict KN light curves. The latest version of \texttt{POSSIS} was used \citep{Bulla2023}, using $10^6$ Monte Carlo quanta (each quanta representing a photon packet) and grid adaptions constructed to simulate the aftermath of a BH-NS merger. The grid describes the composition and architecture of ejecta properties following a merger, setting the stage for evolution of the Monte Carlo quanta that mimic EM radiation produced within the ejecta. A 3D cartesian grid is used, with a resolution of 100 cells per dimension, i.e. a total of $100 \times 100 \times 100 = 10^6$ cells. The ejecta morphology is constructed following the BH-NS grid from  \citet{Kawaguchi2020}, with density distributions
\begin{equation}
\label{rho}
    \rho(r,t, \theta) \propto \begin{cases}
      r^{-3}t^{-3} & 0.025c \leq v \leq 0.1c\\
      \Theta(\theta)r^{-2}t^{-3} & 0.1c \leq v \leq v_{\rm dyn, max}
    \end{cases}  
\end{equation} for the wind and dynamical ejecta respectively. Here, $r$, $t$ and $\Theta(\theta)$ are the radius, time and angular dependence respectively. Whilst spherical symmetry is assumed for the wind ejecta, the dynamical ejecta are largely confined to the merger plane, with an angular dependence of the density distribution described by
\begin{equation}
\label{theta}
    \Theta(\theta) = \frac{1}{1+{\rm exp}[-20(\theta - 1.2[{\rm rad]})]}
\end{equation} where $\theta$ is the angle from the polar axis. The assumed density distribution can be seen in Fig.~\ref{fig:density} for one model in the grid. Ejecta are assumed to undergo homologous expansion throughout: a constant velocity, $v$, for each cell within the grid, with a positive linear relation between $r$ and $v$. Hence ejecta at larger radii travel at higher velocities. The system is thus described in velocity space, as it better represents the expansion of the system (see Fig.~\ref{fig:density}). The wind ejecta are restricted to low velocities and within a range that is fixed for all simulations ($v_{\rm wind, min}=0.025c$ and $v_{\rm wind, max}=0.1c$), while the dynamical ejecta reach higher velocities  ($v_{\rm dyn, min}=0.1c$) with a different extension ($v_{\rm dyn, max}$) for each simulation depending on the value of $v_{\rm dyn}$, equations (\ref{vdyn} and \ref{rho}). 

Values calculated for $M_{\rm dyn}$, $M_{\rm wind}$ and $v_{\rm dyn}$ were used as input to the grid, alongside fixed values for $v_{\rm wind, min}$  and $v_{\rm wind, max}$, equation (\ref{rho}). Fixed values for the electron fractions of the dynamical ejecta ($Y_{\rm e, dyn}$) and wind ejecta ($Y_{\rm e, wind}$) were also used as input: $Y_{\rm e, dyn}$ was assumed to be lanthanide rich material and was assigned a value of 0.1, whilst $Y_{\rm e, wind}$ was assigned a flat distribution between 0.2 and 0.3.  All fixed values were taken from the BH-NS grid in  \citet{Kawaguchi2020}.

% Density figure

\begin{figure*}
	\includegraphics[width=0.9\textwidth]{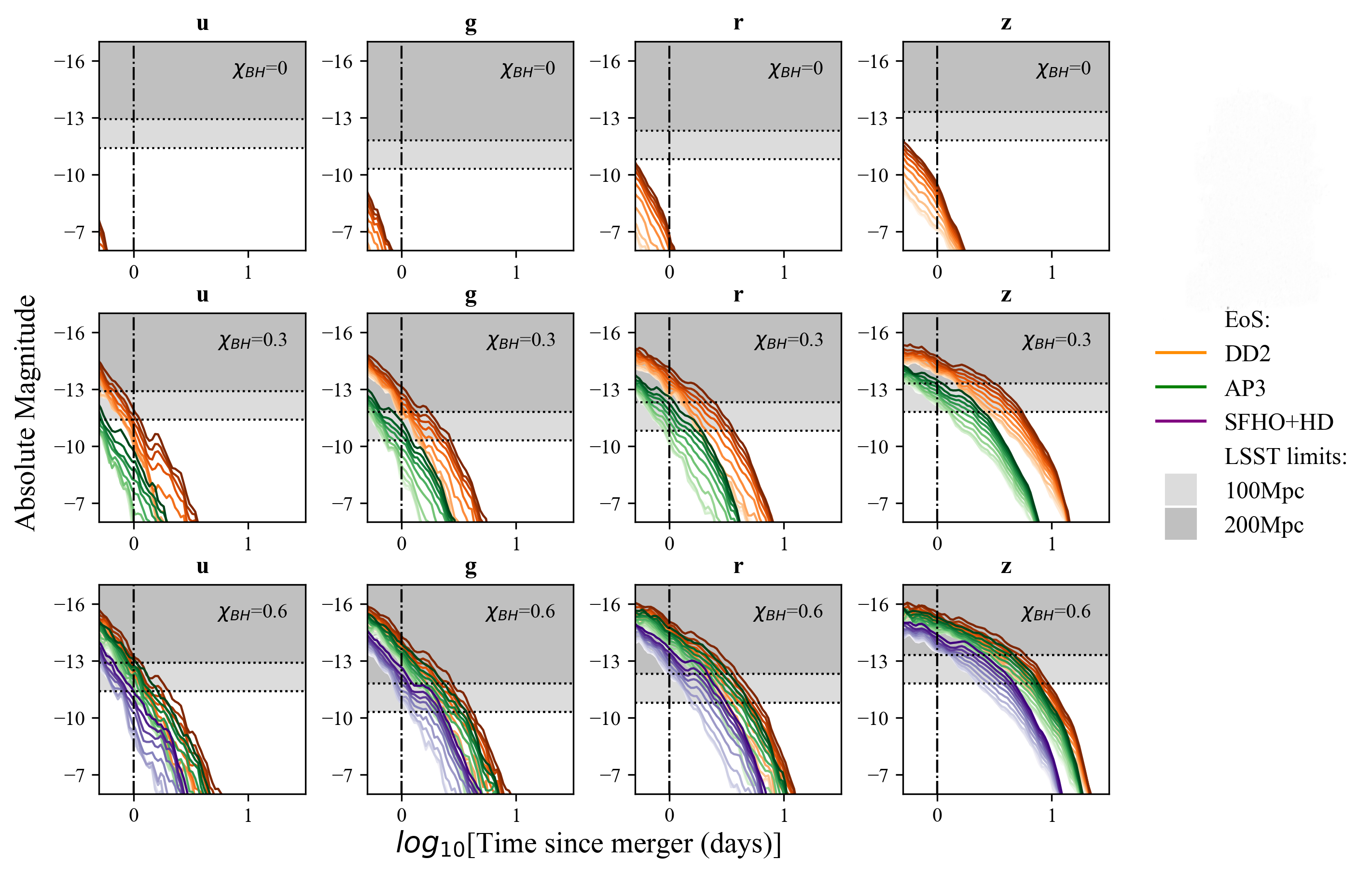}
    \caption{ KN light curves produced by 3 merger systems (1 merger per row) are shown in the u, g, r and z broad-band filters. All merger systems possess $M_{\rm NS} = 1.2M_{\odot}$, $M_{\rm BH} = 6M_{\odot}$, but varying $\chi_{\rm BH}$. Light curves are shown for DD2, AP3 and SFHo+H$\Delta$. Sequential colourmaps for light curves of each EoS indicate observation angle: colours ranging from dark to light (e.g. dark orange to light orange) indicate an observation angle ranging from a polar axis (face-on view) to the merger plane (edge-on view). Vertical lines indicate 1 day post-merger, and horizontal lines indicate LSST limits; magnitudes detectable at 100 and 200 Mpc are coloured in two shades of grey.      
    }
    \label{fig:1.2-6}
\end{figure*}

\subsection{Radiative transfer simulations}
\label{sec:2.4}

Simulations were run in \texttt{POSSIS} for each of the 37 combinations of binary properties and EoS that produced ejecta. Outputs from \texttt{POSSIS} contained information on the flux of the KN, which was used to produce light curves. The flux was provided at 100 logarithmically spaced time intervals from 0.1 to 30 days after the merger. It was provided for 11 observation angles equally spaced in cosine between $\cos\theta_{\rm obs}=0^{\circ}$ (face-on/polar) to $\cos\theta_{\rm obs}=90^{\circ}$ (edge-on/merger plane), for a fixed azimuthal angle phi in the merger plane (exploiting the axial symmetry of the models). The spectral time series from 0.5\footnote{Epochs earlier than 0.5 days are not considered in this work since the opacities adopted in \texttt{POSSIS} prior to this time are likely underestimated, and the brightness of the KN would thus likely be over predicted (see \citet{Tanaka2020} and \citet{Bulla2023} for details).} to 30 days were used to construct KN light curves in the broad-band filters u, g, r and z; these filters encompass the wavelength range of radiation expelled in KNe from early to late times \citep{Metzger2019}.

\section{Results and discussion}

\begin{figure*}	\includegraphics[width=0.8\textwidth]{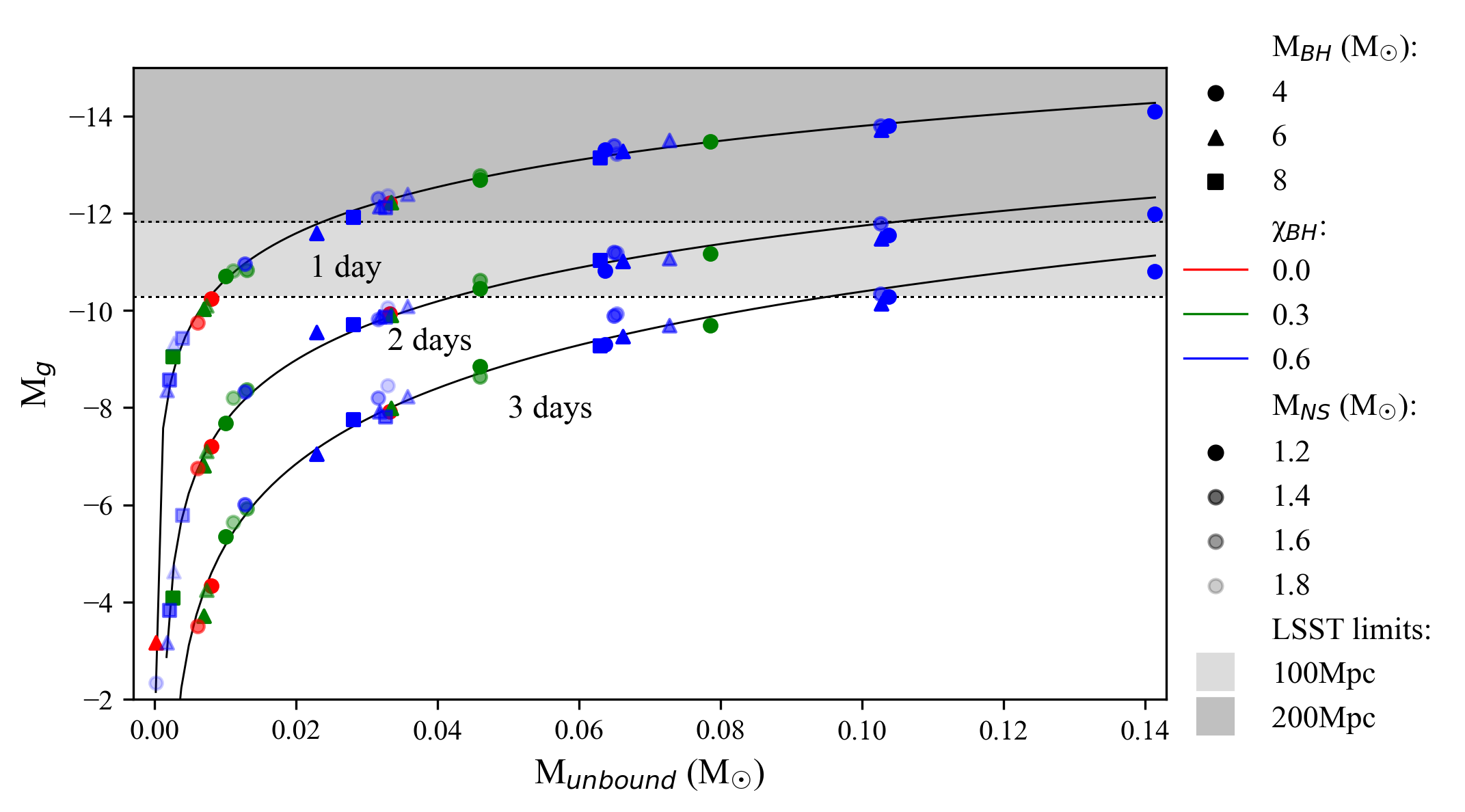}
    \caption{ $M_{\rm g}$ of KNe at 1, 2 and 3 days post-merger as a function of $M_{\rm unbound}$ for all merger systems and EoSs investigated. Each data point at a given epoch represents a merger system and EoS combination. Black solid lines are curves fitted to data at each epoch. Horizontal lines indicate LSST limits in the g band; magnitudes detectable at 100 and 200 Mpc are coloured in two shades of grey. Only datapoints where $M_{\rm g} < 0$ were considered when fitting the curve, since cases where $M_{\rm g} > 0$ correspond to faint and noisy light curves that decrease the accuracy of the fit at detectable magnitudes.
    }
    \label{fig:magvsmtot}
\end{figure*}

The absolute magnitudes of predicted KNe presented in Fig.~\ref{fig:1.2-6}, Fig.~\ref{fig:magvsmtot}, Fig.~\ref{fig:absmagvseos} are overlayed on observational limits of the Vera C. Rubin Observatory's Legacy Survey of Space and Time (LSST) \citep{ivezic2019}, as reported in  \citet{Chase2022}. The limits are specific to each broad-band filter. Limits for 100 and 200 Mpc are shown as this is the distance range where the majority of mergers involving a NS were detected in O3 \citep{Kasliwal2020}. First light from LSST is expected in mid-2024 and its operation to overlap with O5.

\subsection{KN light curves}
\label{sec:3.1}

Predicted KN light curves are analysed for EoSs in the 1-family versus 2-families scenario. 3 merger systems (of the 23 that produced ejecta for at least 1 of the 3 EoSs in question) with the following combination of binary properties are discussed: all with $M_{\rm NS} = 1.2M_{\odot}$, $M_{\rm BH} = 6M_{\odot}$, but varying $\chi_{\rm BH}$ (0, 0.3 and 0.6). These values were selected for discussion as they all fall within uncertainty ranges of the binary properties identified from the GW signal of BH-NS merger GW200115 detected during O3 in 2020 \citep{AbbottR2021}; no KN was detected \citep{Anand2021}. The plots are displayed from 0.5 to 30 days with a logarithmic time scale. Light curves for the remaining 20 merger systems that produced ejecta and thus a KN are shown in Appendix \ref{appendix:appendixA}. 

Fig.~\ref{fig:1.2-6} shows that a stiffer EoS increases the likelihood of ejecta being produced by the merger and therefore, produces a brighter KN; these results are consistent with findings in  \citet{Barbieri2020}. For this reason, the EoSs from the 1-family scenario (DD2, AP3) are more likely to produce KNe in all broad-band filters than the EoS from the 2-families scenario (SFHo+H$\Delta$). 

A relation between $\chi_{\rm BH}$ and the KN can be deduced from Fig.~\ref{fig:1.2-6}; higher spins result in brighter KNe for a given EoS. This is consistent with findings in  \citet{DiClemente2022}, where higher spins are shown to eject more mass during a merger. It is also consistent with the condition required for the production of merger ejecta, $d_{\rm tidal} > R_{\rm ISCO}$ \citep{Foucart2012}; for constant values of  $M_{\rm NS}$, $M_{\rm BH}$ and $R_{\rm NS}$ (i.e. constant $d_{\rm tidal}$), the condition is more easily met for higher $\chi_{\rm BH}$.

For all light curves displayed in Fig.~\ref{fig:1.2-6}, a relation between the brightness of the KN and the observation angle can be identified; an observer viewing the system face-on (along the polar axis) sees a brighter KN. This is due to the angular dependence of $M_{\rm dyn}$ in BH-NS mergers, expressed in equation (\ref{rho}) and seen in Fig.~\ref{fig:density}. The asymmetry of the dynamical ejecta results in larger quantities of lanthanide-rich material forming about the merger plane than closer to the polar axis. This lanthanide rich material is associated with high opacities which slows the EM radiation traversing it, resulting in fainter light curves observed in the merger plane.

Once LSST becomes operational, light curves in Fig.~\ref{fig:1.2-6} could be used to aid in identifying whether the 1-family or 2-families scenario holds true in nature, and in turn constrain the EoS of NS matter. By first identifying the binary properties of a BH-NS merger ($M_{\rm NS}$, $M_{\rm BH}$, $\chi_{\rm BH}$), calculating the distance to the merger, and constraining the observation angle (all from the GW signal that precedes the KN), the presence or absence of a KN signal would indicate which EoS best describes the NS. 
Notice anyway that GW observational data from LIGO-Virgo  provide a fairly accurate measurement of the chirp mass \citep{LIGO2015,Virgo2015}, but not an equally accurate measurement of the spin and individual masses of the components of a merger. Also, when performing the data analysis, the values of the masses and spins turn out to be strongly correlated. In a previous paper we have analyzed the impact of these observational uncertainties and we have shown how they affect the probability of observing a KN signal \citep{DiClemente2022}. We have shown that, if the "simple" analysis not accounting of the observational uncertainties indicates a strong KN signal (when compared with the sensitivity of LSST), the "complete" analysis shows a high probability that the KN signal can indeed be observed. Similarly, if the signal is estimated to be very weak, the complete analysis suggests a very low probability of detection. In other terms, the present observational errors associated with the GW signal do not spoil the possibility of reaching a conclusion concerning the EoS if the signal is strong.

\subsection{Relation between absolute magnitude of a KN and the mass of unbound material ejected from the BH-NS merger}
\label{sec:3.2}

Further analysis is carried out on the absolute magnitude of KNe in the $g$ band ($M_{\rm g}$): $M_{\rm g}$ is plotted as a function of the gravitationally unbound ejected mass from the merger ($M_{\rm unbound} = M_{\rm dyn} + M_{\rm wind}$). $M_{\rm g}$ is averaged over all observing angles. Results for all 37 combinations of merger system and EoS that produced ejecta are combined on a single plot (Fig.~\ref{fig:magvsmtot}).

As expected, a positive correlation is identified between $M_{\rm g}$ and $M_{\rm unbound}$ (a similar correlation exists in other broad-band filters); an increase in ejecta mass leading to an increase in r-process material surrounding the remnant, resulting in brighter KNe. The correlation was found to be a tight logarithmic fit, consistent across all epochs studied between 1 and 6 days (see Fig.~\ref{fig:magvsmtot} for results between 1 and 3 days), albeit with increasing scatter with time. Curves were fitted to data at 1, 2 and 3 days post-merger; KNe will likely be too faint for detection at later times according to the LSST limits seen in Fig.~\ref{fig:magvsmtot}, so no fits were produced at later times. The following logarithmic fits were found

\begin{equation}
\label{Mg1day}
    M_{\rm g}^{\rm 1d} = -3.150 \log_{10} \left(M_{\rm unbound}-0.0002 \right)-16.95 
\end{equation} 

\begin{equation}
\label{Mg2day}
    M_{\rm g}^{\rm 2d} = -3.833 \log_{10} \left(M_{\rm unbound}-0.0012 \right)-15.60  
\end{equation} 

\begin{equation}
\label{Mg3day}
    M_{\rm g}^{\rm 3d} = -4.855 \log_{10} \left(M_{\rm unbound}-0.0018 \right)-15.60 
\end{equation} where $M_{\rm g}^{\rm 1d}$, $M_{\rm g}^{\rm 2d}$,  $M_{\rm g}^{\rm 3d}$ are $M_{\rm g}$ at 1, 2 and 3 days post-merger respectively, and $M_{\rm unbound}$ is the total unbound ejecta mass from the merger in solar masses. 

These logarithmic fits are useful to rapidly calculate $M_{\rm unbound}$ from the observed magnitude $M_g$. In this way, it is possible to check the compatibility of an observed KN signal with a specific EoS without having to run a time-consuming RT simulation such as \texttt{POSSIS}.

\begin{figure*}
    \includegraphics[width=0.9\textwidth]{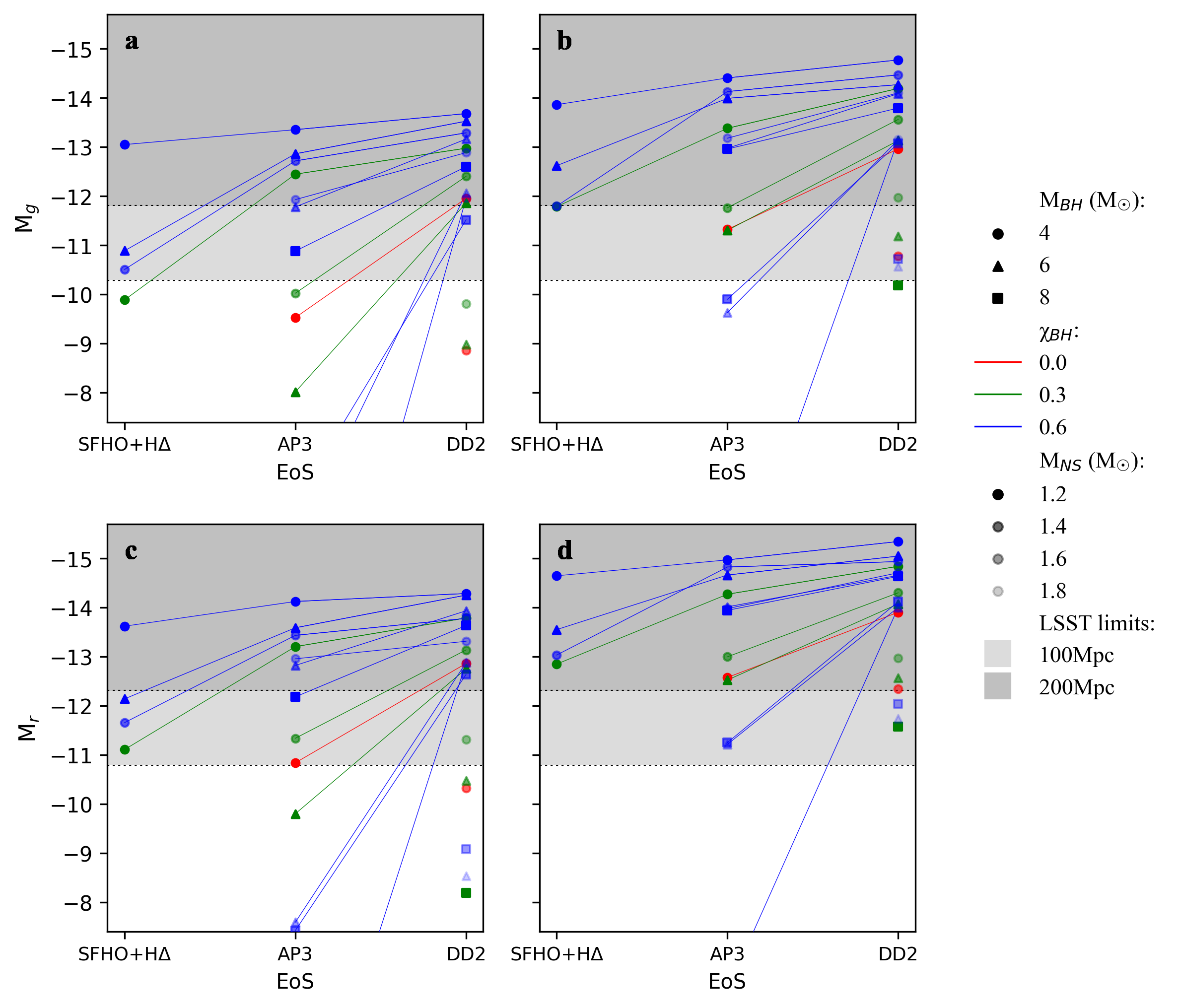}
    \caption{ Absolute magnitude in the $g$ and $r$ bands (top and bottom rows respectively) of KNe at 1 day post merger as a function of EoS for various merger systems. Panels a and c show magnitudes seen in the merger plane ($\cos\theta_{\rm obs}=90^{\circ}$) whilst panel b and d show those from the polar axis ($\cos\theta_{\rm obs}=0^{\circ}$).The stiffness of the EoS increase along the x-axis of each plot. Solid lines join merger systems with identical binary properties for each EoS. Horizontal lines indicate relevant LSST limits in the $g$ and $r$ bands; magnitudes detectable at 100 and 200 Mpc are coloured in two shades of grey.       
    }
    \label{fig:absmagvseos}
\end{figure*}

\begin{figure*}
    \includegraphics[width=0.45\textwidth]{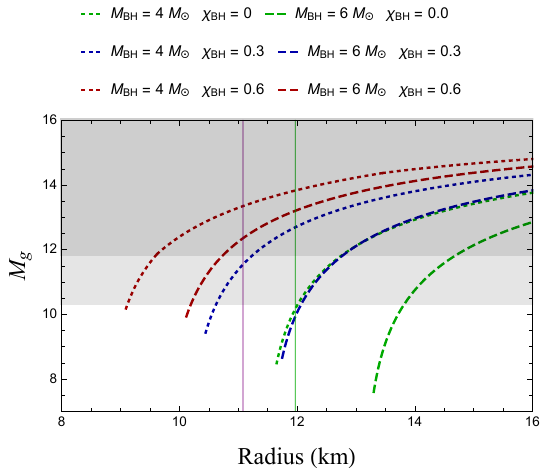}
    \includegraphics[width=0.45\textwidth]{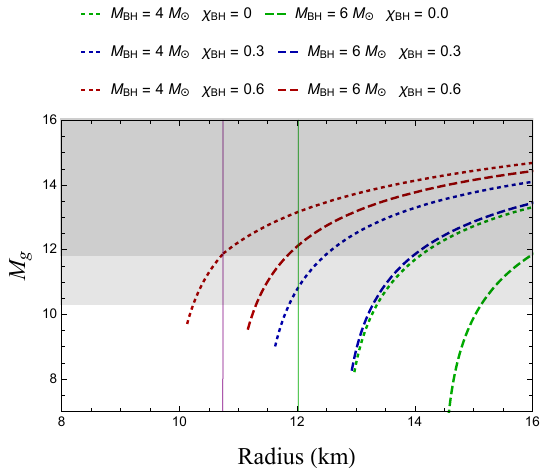}
    \caption{\revised{Absolute g-band magnitudes at 1 day after the merger as a function of NS radius for systems with $M_{\rm NS}=1.2\,M_\odot$ (left panel) and $M_{\rm NS}=1.4\,M_\odot$ (right panel) and varying BH masses and spins. Curves are computed by estimating the total ejected mass $M_{\rm tot}$ from Equation~\ref{Mtot} \citep{Foucart2018} and its connection with the g-band magnitude at 1 day from Equation~\ref{Mg1day}. The purple and green vertical lines mark the NS radii for the SFHo+H$\Delta$ (2-family) and the AP3 (1-family) EoSs. Magnitudes detectable by VRO/LSST at 100 and 200 Mpc are coloured in two shades of grey.}       
    }
    \label{fig:mag_radii}
\end{figure*}

\subsection{Relation between absolute magnitude of a KN and EoS }
\label{sec:3.3}

We further investigate the relation between the absolute magnitude of KNe in both the $g$ ($M_{\rm g}$) and $r$ bands ($M_{\rm r}$) and the EoS describing the NS (Fig.~\ref{fig:absmagvseos}); it is again shown that stiffer EoSs produce a brighter KN for a given merger system. This relation is true for all EoSs (DD2, AP3 and SFHo+H$\Delta$) across every merger system investigated, again consistent with findings in  \citet{Barbieri2020}. 

As with previous plots discussed, Fig.~\ref{fig:absmagvseos} could also be used to constrain the EoS of NS matter by aiding in identifying which of the two scenarios hold true in nature. By first identifying the binary properties of a BH-NS merger from a detected GW, the $M_{\rm g}$ or $M_{\rm r}$ of the KN would be predicted using these plots for the identified set of binary properties, for each EoS. The $M_{\rm g}$/ $M_{\rm r}$ measured from the KN signal following the GW would be compared to $M_{\rm g}$/ $M_{\rm r}$ values predicted here for each EoS; EoSs predicting vastly different values to those measured would be ruled out. This would in turn constrain the EoS of NS matter. 

For instance, Fig.~\ref{fig:absmagvseos} shows that an EoS from the two families scenario (SFHo+H$\Delta$) is much less likely to produce a KN; a KN was only produced in 4 of the merger systems investigated. These merger systems comprised of low mass BHs ($\leq 6M_{\odot}$), low mass NSs ($\leq 1.4M_{\odot}$) and high $\chi_{\rm BH}$ ($\geq 0.3$) (see Table~\ref{tab:table_2} for details). On the contrary, strong KN signals are seen at larger BH and NS masses and lower $\chi_{\rm BH}$ in the 1-family scenario (DD2 and AP3). Additionally, a strong KN signal is seen in  the 1-family scenario for merger systems involving a non-rotating BH ($\chi_{\rm BH} = 0$) when $M_{\rm BH} = 4M_{\odot}$ and $M_{\rm NS} = 1.2M_{\odot}$; this result is consistent with findings in  \citet{DiClemente2022}. For example, the detection of a KN by LSST, from a merger involving  a non-rotating BH of mass $M_{\rm BH}= 4M_{\odot}$ and a NS of mass $M_{\rm NS}= 1.2M_{\odot}$ at a distance $\leq 100$ Mpc, would indicate that a NS from the 1-family scenario was responsible for its formation (since a KN should not be present for the 2-families scenario). \revised{The constraining power can be further improved with the Nancy Grace Roman Space Telescope \citep{Spergel2015}, which compared to LSST will allow to reach deeper magnitudes in the near infrared ($\sim2.5$ magnitude in the $i-$band, see \citealt{Chase2022}) and extend to even longer wavelengths \citep{Andreoni:2023xlv}.}

\section{Conclusions and limitations}

KN light curves and their absolute magnitudes were investigated for 36 unique BH-NS merger systems and 3 EoSs. Analysis showed that stiffer EoSs result in brighter KNe, consistent with simulations in  \citet{DiClemente2022} and  \citet{Barbieri2020}. This provides a concrete way to test the two-families scenario, since the hadronic family in that scheme corresponds to a very soft EoS and therefore to a very weak KN signal. We have shown that the KN signal produced in the 2-families scenario is so weak that it should be possible to distinguish between the two scenarios even by using the GW data collected in O4 or O5 and the sensitivity of \revised{future observational facilities as e.g. VRO/LSST and the Nancy Grace Roman Space Telescope}. A caveat is in order: we have excluded the possibility of producing a KN signal from a BH-QS merger, because of the results of the simulation in \citet{Kluzniak2002}. Notice anyway that in that simulation the BH was not rotating and we have shown that the amount of mass ejected is strongly reduced in that case.

\revised{The SFHo+H$\Delta$ (2-family) and the AP3 (1-family) EoSs are characterized by differences of $\sim$1 and $\sim$1.5 km for 1.2 and 1.4\,$M_\odot$ NSs, respectively. As shown in Fig.~\ref{fig:mag_radii}, these correspond to differences of $\gtrsim$1 mag in g-band magnitudes for the systems considered in this work. These differences are large enough to be distinguished with kilonova observations in the future. However, EoSs where the differences between a 1-family and a 2-family scenario are smaller would be more challenging to tell apart with kilonovae. For instance, two EoSs with a difference in radius of $\lesssim500$m for a 1.2 $M_\odot$ NS can lead to g-band light curves that differ of $\lesssim0.5$mag at 1 day and that would be hard to distinguish with kilonova observations.}

In principle, one can hope to distinguish also between various EoSs belonging to the 1-family scenario.
This possibility is highly dependent on the ability to constrain binary properties (especially $\chi_{\rm BH}$) from the preceding GW signal. Due to the limitations of current technologies, constraining the EoS from Fig.~\ref{fig:1.2-6} is still difficult, unless the signal is expected to be particularly strong. With improved sensitivities of the advanced LIGO \citep{LIGO2015}, advanced Virgo \citep{Virgo2015} and KAGRA \citep{KAGRA2019} interferometers in future runs and the construction of new interferometers such as LIGO-India, the ability to identify binary properties will be greatly improved, enabling much tighter constraints on the EoS of NS matter.

By combining results from all merger systems and EoSs a positive logarithmic correlation was found between the $M_{\rm g}$ and $M_{\rm unbound}$; more ejecta mass leading to an increase in r-process rich material surrounding the remnant, resulting in brighter KNe. The correlation was consistent across all epochs studied between 1 and 6 days post-merger, albeit with an increase in scatter with time. Logarithmic fits were provided at 1, 2 and 3 days and they can be used to significantly simplify the analysis of future events.

The main limitation of this work lies in the fitting formulae used to calculate merger ejecta properties, which are known to introduce biases \citep{Raaijmakers2021_2}; assumptions were made in calculations of $M_{\rm dyn, max}$ and $M_{\rm wind}$. \revised{However, an average uncertainty of $\sim$15\% on the total ejecta mass, as estimated in \cite{Foucart2018}, translates into an error of only $\sim$0.2\,mag in the g-band light curve at 1 day according to Equation~\ref{Mg1day} we derived in our work}. Limited work exists in the literature that explores the properties of ejecta mass from BH-NS mergers, making accurate assumptions in the fitting formulae challenging. The value of $f_1$ used in the calculation of $M_{\rm dyn \ max}$ was chosen based on simulations in  \citet{Foucart2019}, which focus on the \revised{near equal-mass regime} (where $M_{\rm NS} \simeq M_{\rm BH}$) of BH-NS mergers, unlike in this paper. This may have subsequently resulted in errors in predictions of KN brightness. Similarly, the value of $f_2$ used to calculate $M_{\rm wind}$ is unlikely to be precise; the multiple mechanisms simultaneously determining the evolution of the disk ejecta are currently poorly understood, making accurate predictions of $M_{\rm wind}$ difficult. Future work will likely benefit from rapidly evolving interest and research in BH-NS mergers ahead of their anticipated detection in O4 and future runs.

\section*{Acknowledgements}

We thank the anonymous reviewer for helpful comments that improved this paper. This work was supported by the European Union’s Horizon 2020 Programme under the AHEAD2020 project (grant agreement n. 871158) and by the National Science Foundation under Grant No. PHY-1430152 (JINA Center for the Evolution of the Elements). The simulations were performed on resources provided by the Swedish National Infrastructure for Computing (SNIC) at Kebnekaise partially funded by the Swedish Research Council through grant agreement no. 2018-05973.

\section*{Data Availability}

The simulations performed in this study will be made publicly available at \url{https://github.com/mbulla/kilonova_models}.

\bibliographystyle{mnras}
\bibliography{example} 
\newpage

\appendix

\section{KN light curves}
\label{appendix:appendixA}
Light curve plots for the remaining simulated merger systems that produced ejecta and thus a KN are shown below.

\begin{figure*}
	\includegraphics[width=0.8\textwidth]{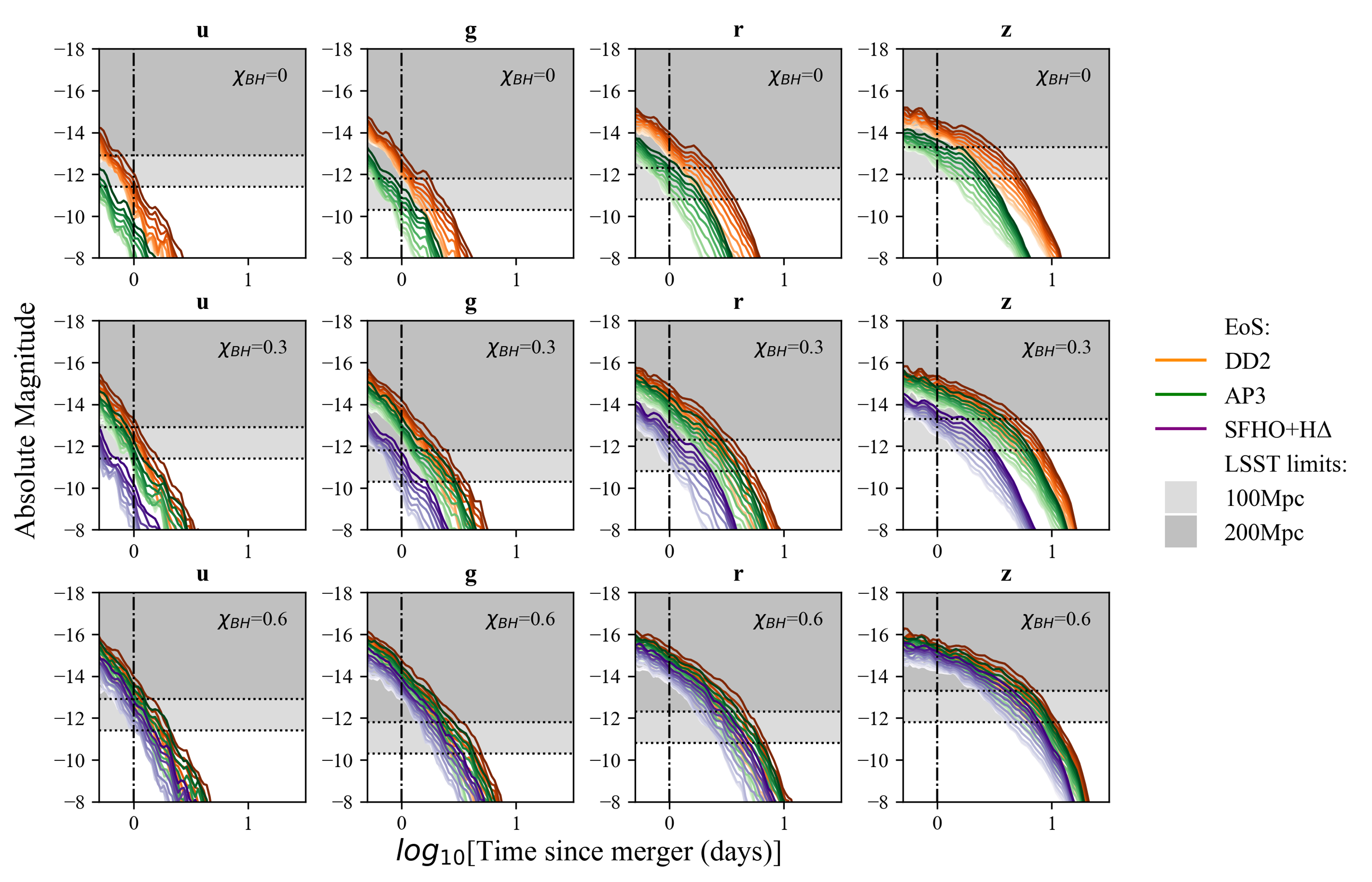}
    \caption{KN light curves produced by the merger systems $M_{\rm NS}=1.2M_{\odot}$, $M_{\rm BH}=4M_{\odot}$, $\chi_{\rm BH}= 0$, $0.3$ and $0.6$.     
    }
    \label{fig:1.2-4}
\end{figure*}

\begin{figure*}
	\includegraphics[width=0.8\textwidth]{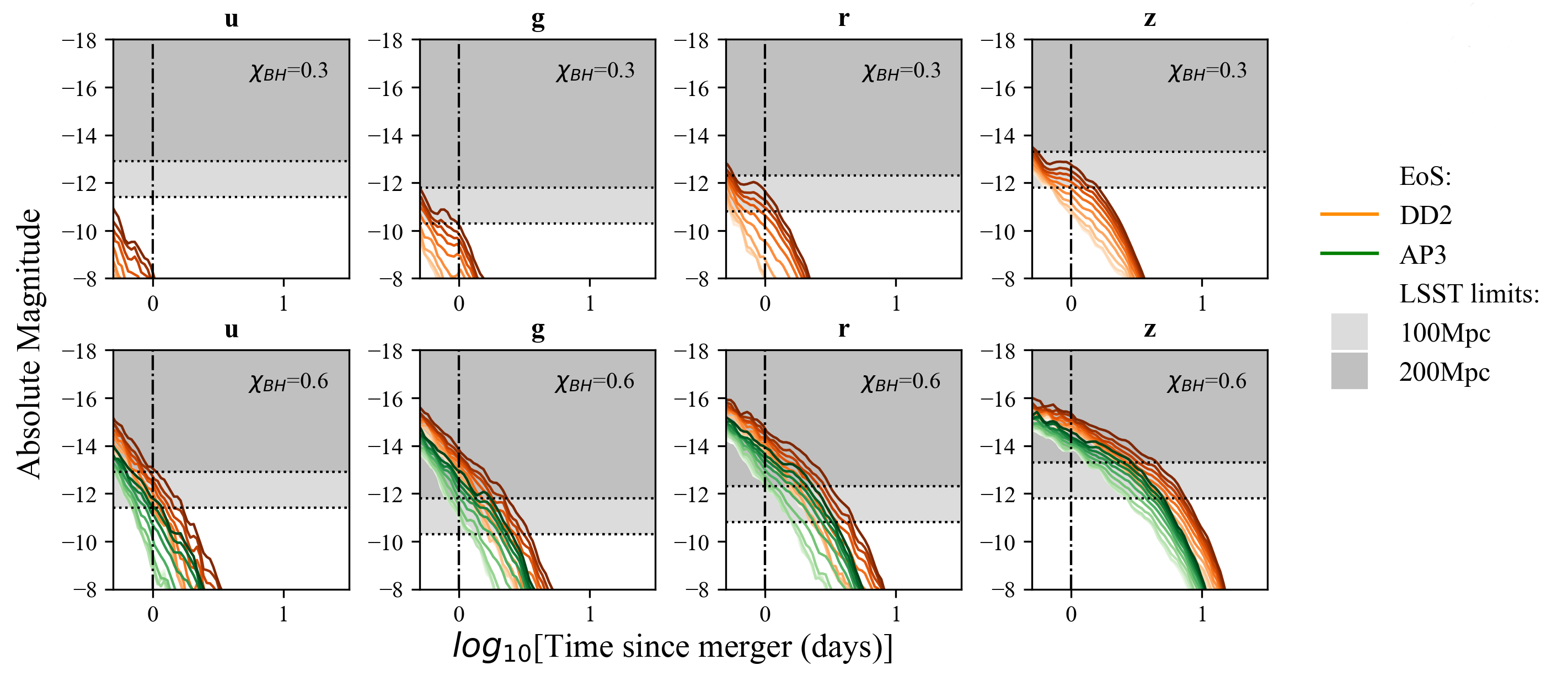}
    \caption{KN light curves produced by the merger systems $M_{\rm NS}=1.2M_{\odot}$, $M_{\rm BH}=8M_{\odot}$, $\chi_{\rm BH}= 0.3$ and $0.6$.     
    }
    \label{fig:1.2-8}
\end{figure*}

\begin{figure*}
	\includegraphics[width=0.8\textwidth]{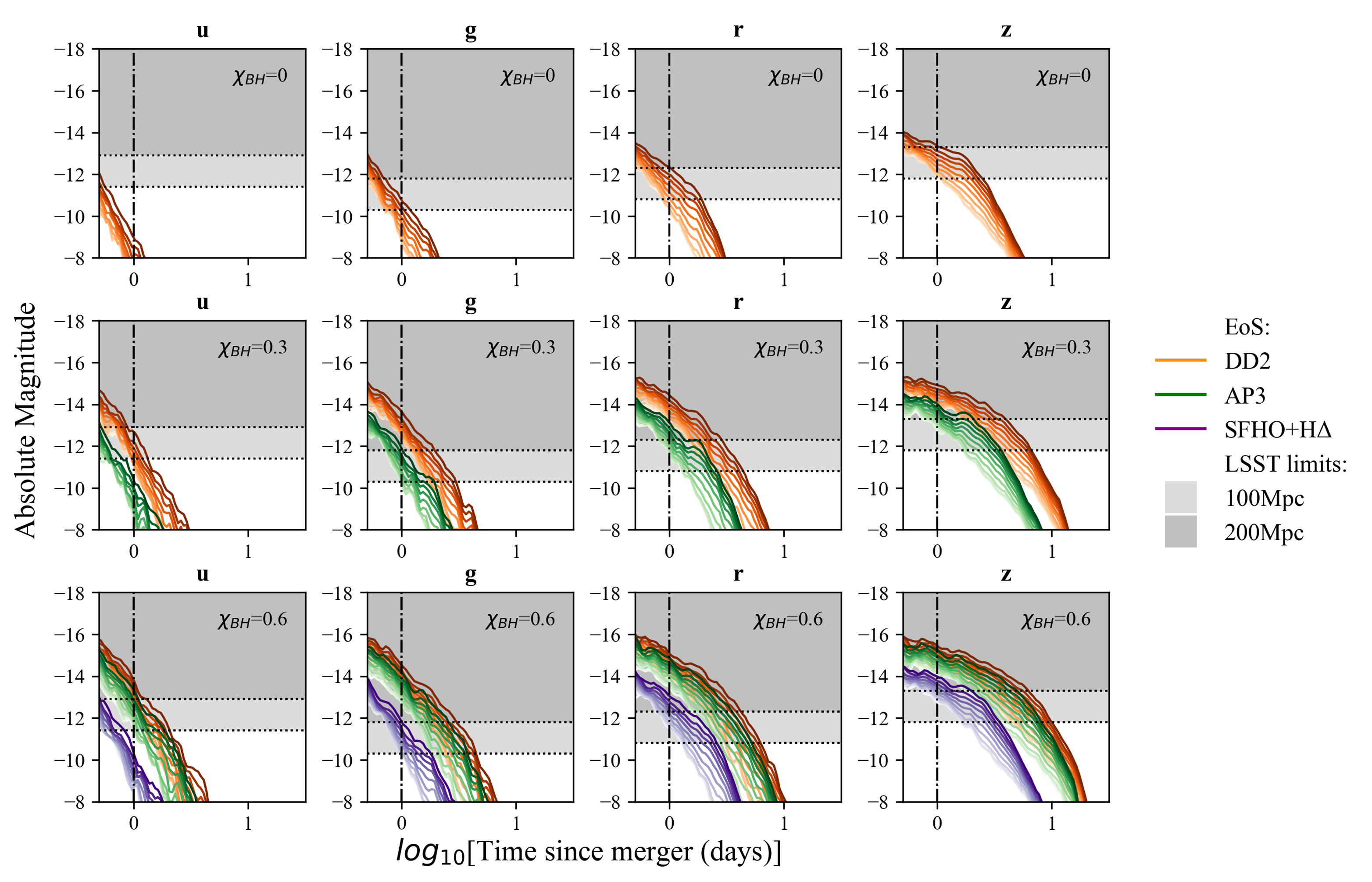}
    \caption{KN light curves produced by the merger systems $M_{\rm NS}=1.4M_{\odot}$, $M_{\rm BH}=4M_{\odot}$, $\chi_{\rm BH}= 0$, $0.3$ and $0.6$.     
    }
    \label{fig:1.4-4}
\end{figure*}

\begin{figure*}
	\includegraphics[width=0.8\textwidth]{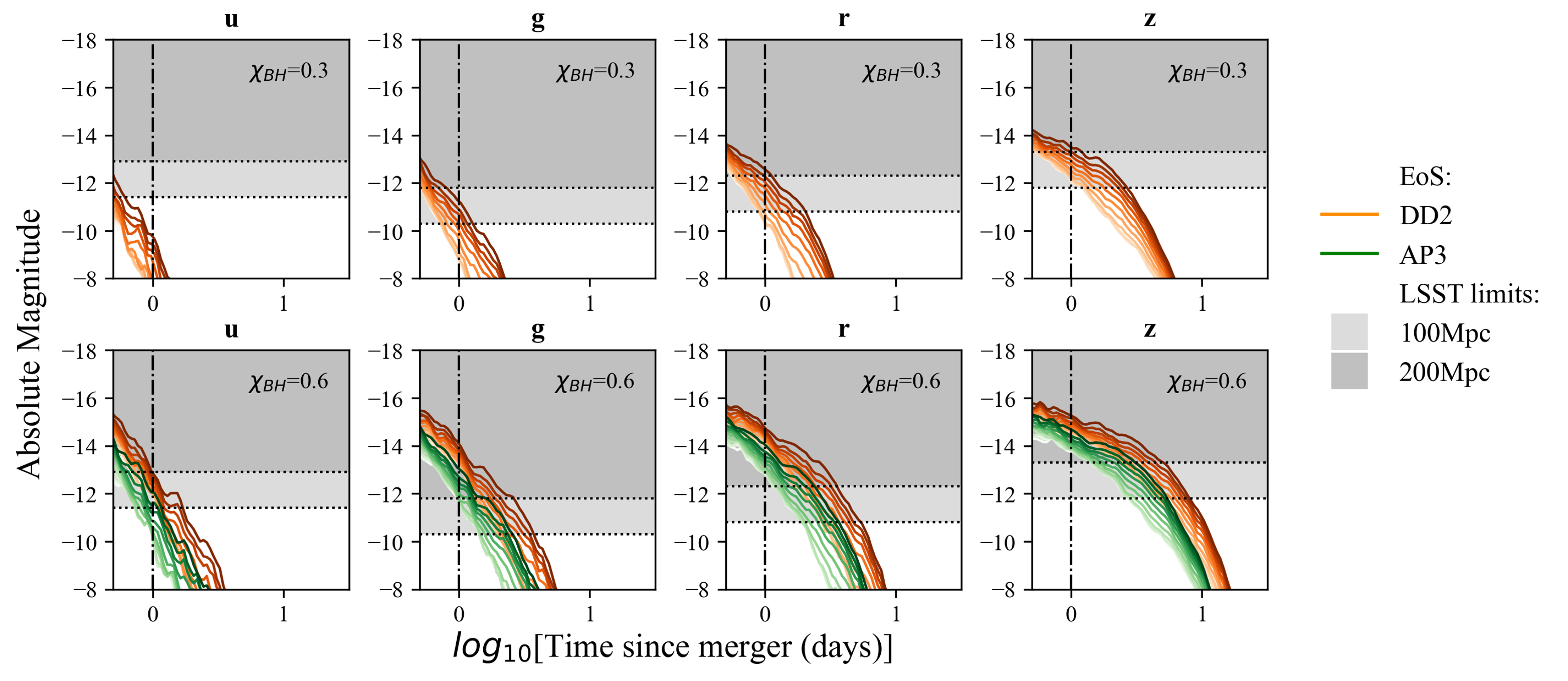}
    \caption{KN light curves produced by the merger systems $M_{\rm NS}=1.4M_{\odot}$, $M_{\rm BH}=6M_{\odot}$, $\chi_{\rm BH}= 0.3$ and $0.6$.     
    }
    \label{fig:1.4-6}
\end{figure*}

\begin{figure*}
	\includegraphics[width=0.8\textwidth]{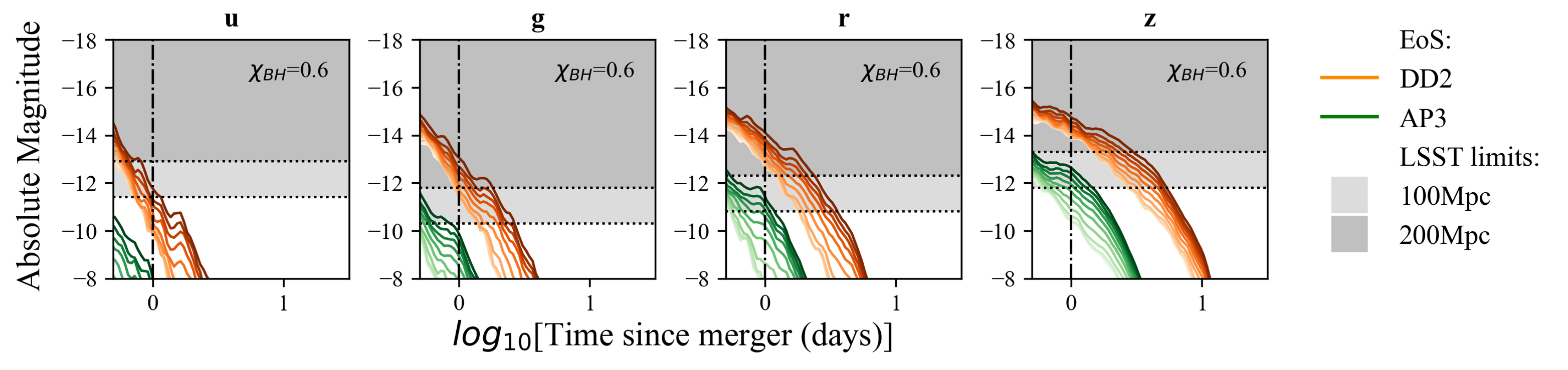}
    \caption{KN light curves produced by the merger systems $M_{\rm NS}=1.4M_{\odot}$, $M_{\rm BH}=8M_{\odot}$, $\chi_{\rm BH}= 0.6$.     
    }
    \label{fig:1.4-8}
\end{figure*}

\begin{figure*}
	\includegraphics[width=0.8\textwidth]{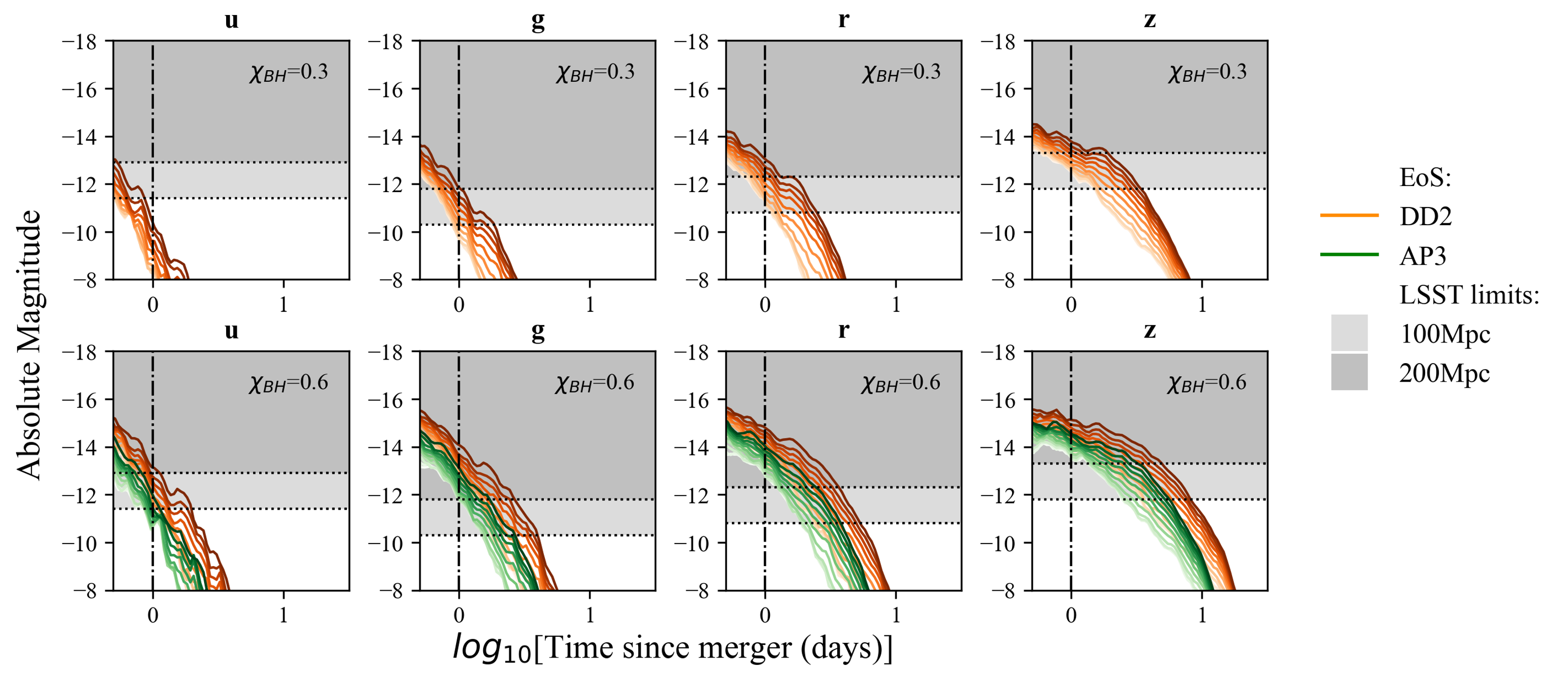}
    \caption{KN light curves produced by the merger systems $M_{\rm NS}=1.6M_{\odot}$, $M_{\rm BH}=4M_{\odot}$, $\chi_{\rm BH}= 0.3$ and $0.6$.     
    }
    \label{fig:1.6-4}
\end{figure*}

\begin{figure*}
	\includegraphics[width=0.8\textwidth]{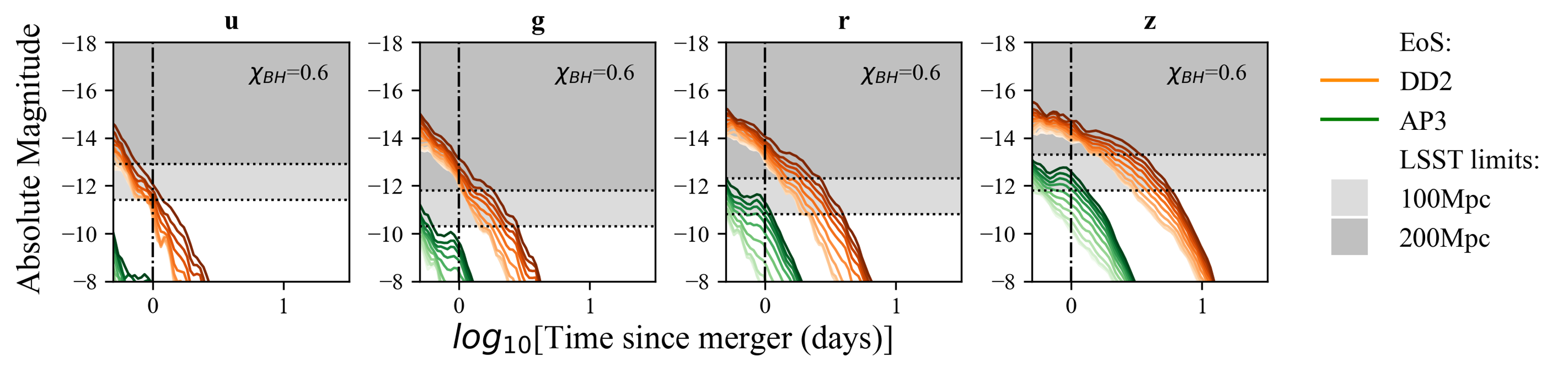}
    \caption{KN light curves produced by the merger systems $M_{\rm NS}=1.6M_{\odot}$, $M_{\rm BH}=6M_{\odot}$, $\chi_{\rm BH}= 0.6$.     
    }
    \label{fig:1.6-6}
\end{figure*}

\begin{figure*}
	\includegraphics[width=0.8\textwidth]{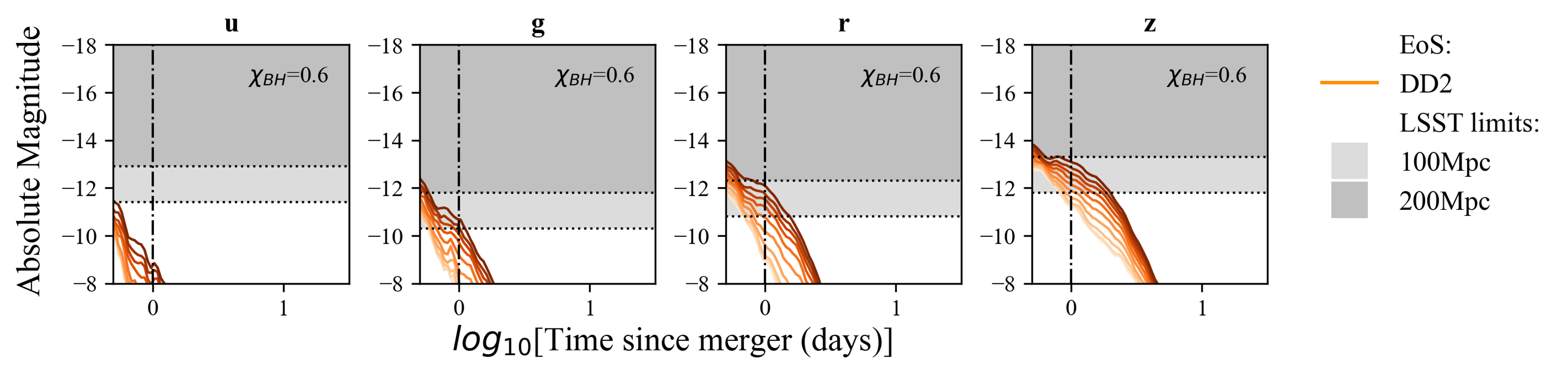}
    \caption{KN light curves produced by the merger systems $M_{\rm NS}=1.6M_{\odot}$, $M_{\rm BH}=8M_{\odot}$, $\chi_{\rm BH}= 0.6$.     
    }
    \label{fig:1.6-8}
\end{figure*}

\begin{figure*}
	\includegraphics[width=0.8\textwidth]{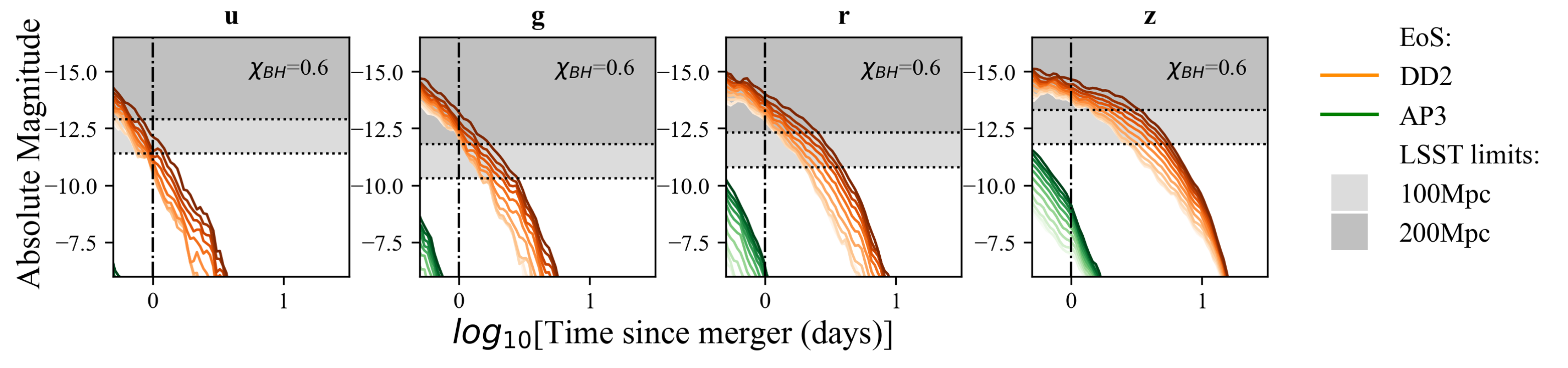}
    \caption{KN light curves produced by the merger systems $M_{\rm NS}=1.8M_{\odot}$, $M_{\rm BH}=4M_{\odot}$, $\chi_{\rm BH}= 0.6$.     
    }
    \label{fig:1.8-4}
\end{figure*}

\begin{figure*}
	\includegraphics[width=0.8\textwidth]{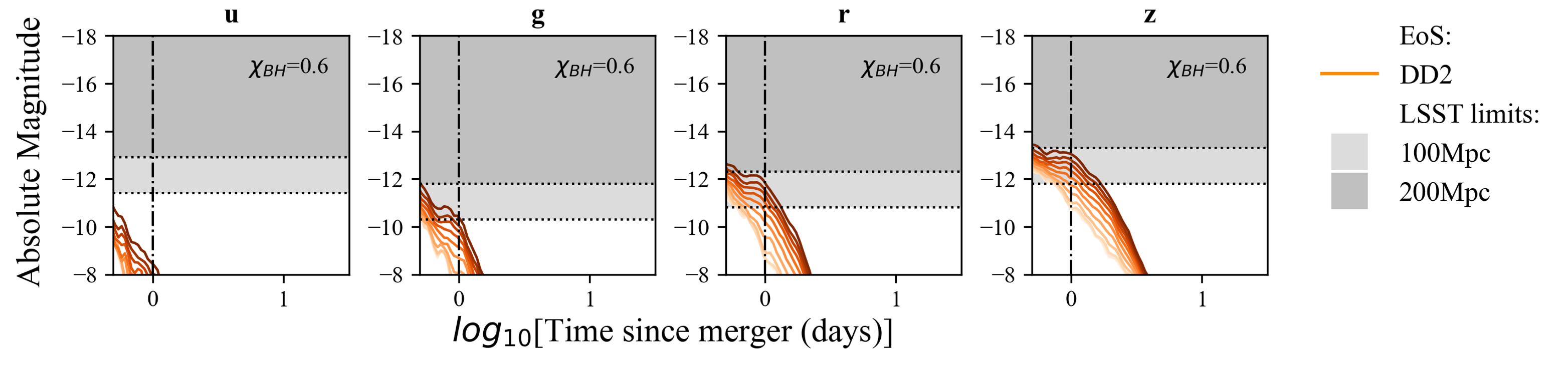}
    \caption{KN light curves produced by the merger systems $M_{\rm NS}=1.8M_{\odot}$, $M_{\rm BH}=6M_{\odot}$, $\chi_{\rm BH}= 0.6$.     
    }
    \label{fig:1.8-6}
\end{figure*}

\bsp	
\label{lastpage}
\end{document}